\documentclass[review,3p,11pt]{elsarticle}
\usepackage[english]{babel}
\usepackage{amsmath}
\usepackage{graphicx}
\usepackage{hyperref}
\usepackage{soul}
\usepackage{lineno}

\journal{Physica. A}

\begin{document}
\title{Crime prediction through urban metrics and statistical learning}
 
\author[1]{Luiz G. A. Alves}
\ead{lgaalves@usp.br} 
\author[2]{Haroldo V. Ribeiro}
\author[1]{Francisco A. Rodrigues}

\address[1]{Institute of Mathematics and Computer Science, University of S\~ao Paulo, S\~ao Carlos, SP, Brazil}
\address[2]{Departamento de F\'isica, Universidade Estadual de Maring\'a, Maring\'a, PR, Brazil}

\begin{abstract}
Understanding the causes of crime is a longstanding issue in researcher's agenda. While it is a hard task to extract causality from data, several linear models have been proposed to predict crime through the existing correlations between crime and urban metrics. However, because of non-Gaussian distributions and multicollinearity in urban indicators, it is common to find controversial conclusions about the influence of some urban indicators on crime. Machine learning ensemble-based algorithms can handle well such problems. Here, we use a random forest regressor to predict crime and quantify the influence of urban indicators on homicides. Our approach can have up to $97\%$ of accuracy on crime prediction, and the importance of urban indicators is ranked and clustered in groups of equal influence, which are robust under slightly changes in the data sample analyzed. Our results determine the rank of importance of urban indicators to predict crime, unveiling that unemployment and illiteracy are the most important variables for describing homicides in Brazilian cities. We further believe that our approach helps in producing more robust conclusions regarding the effects of urban indicators on crime, having potential applications for guiding public policies for crime control. 
\end{abstract}
\maketitle

\section*{Highlights}
\begin{itemize}
\item Predictive analysis is applied to crime data;
\item Correlation between crime and urban metrics;
\item Quantification of the importance of urban metrics in predicting crime.
\end{itemize}

\section{Introduction}
Social phenomena increasingly attract the attention of physicists driven by successful application of methods from statistical physics for modeling and describing several social systems, including the collective phenomena emerging from the interactions of individuals~\cite{castellano2009statistical}, the spread of ideas in social networks~\cite{pentland2014social}, epidemic spreading~\cite{pastor2015epidemic}, criminal activity~\cite{d2015statistical}, political corruption~\cite{ribeiro2018dynamical}, vaccination strategies~\cite{wang2016statistical}, and human cooperation ~\cite{perc2017statistical}. In the particular case of crime, this interest trace back to works of Quetelet, who coined the term ``social physics'' in the 19th century~\cite{quetlet1869physique}. On the one hand, traditional physics methods have proved to be useful in understanding phenomena outside conventional physics~\cite{galam2012sociophysics,conte2012manifesto}. On the other hand, recently, several problems from physics have been addressed through the lenses of machine learning methods, including topics related to phases of matter~\cite{carrasquilla2017machine}, quantum many-body problem~\cite{carleo2017solving}, phase transitions~\cite{van2017learning}, phases of strongly correlated fermions~\cite{ch2017machine}, among others. As physicists had added these new tools to the box~\cite{zdeborova2017machine}, naturally, social physics problems could also be addressed using such ideas. In particular, in this work, we are interested in understanding the relationships between crime and urban metrics by using statistical learning.

Crime and violence are ubiquitous in society. Throughout history, organized societies have tried to prevent crime following several approaches~\cite{gordon2009crime}. In this context, understanding the features associated with crime is essential for achieving effective policies against these illegal activities. Studies have linked crime with several factors, including psychological traits~\cite{kamaluddin2015linking,gottfredson1990general}, environmental conditions~\cite{gamble2012temperature,hsiang2013quantifying}, spatial patterns\cite{short2008statistical,alves2015spatial,d2015statistical}, and social and economic indicators~\cite{becker1968crime,ehrlich1973deterrent,wilson1982broken,glaeser1996crime}. However, it is easy to find controversial explanations for the causes of crimes~\cite{gordon2010random}. Methodological problems in data aggregation and selection~\cite{levitt2001alternative,spelman2008specifying}, errors related to data reporting~\cite{maltz2002note}, and wrong statistical hypothesis~\cite{gordon2010random} are just a few issues which can lead to misleading conclusions. 

A significant fraction of the literature on statistical analysis in criminology tries to relate the number of a particular crime (\textit{e.g.} robbery) with explicative variables such as unemployment~\cite{raphael2001identifying} and income~\cite{kelly2000inequality}. In general, these analyses are carried out by using ordinary-least-squares (OLS) linear regressions~\cite{alves2015scale}. These standard linear models usually assume that the predictors have weak exogeneity (error-free variables), linearity, constant variance (homoscedasticity), normal residual distribution, and lack of multicollinearity. However, when trying to model crime, several of these assumptions are, often, not satisfied. When these hypotheses do not hold, conclusions about the factors affecting crime are likely to be misconceptions.

Recently, researchers have promoted an impressive progress on the analysis of cities, where one of the main findings is that the relationship between urban metrics and population size is not linear, but it is well described by a power-law function~\cite{bettencourt2007growth,bettencourt2010urban,alves2013scaling,alves2013distance,alves2014empirical,hanley2016rural,leitao2016scaling}. Crime indicators scale as a superlinear function of the population size of cities~\cite{bettencourt2010urban,alves2013distance,alves2015scale}. Other indicators (commonly used as predictors in linear regression models for crime forecasting) also exhibit power-law behavior with population size. These metrics are categorized into sub-linear (\textit{e.g.} family income~\cite{alves2013distance}), linear (\textit{e.g.} sanitation~\cite{alves2013distance}), and super-liner (\textit{e.g.} GDP~\cite{bettencourt2007growth,alves2013distance,alves2015scale,alves2017role}), depending on the power-law exponent characterizing the allometric relationship with the population size~\cite{bettencourt2007growth}. In addition, the relationships between crime and population size as well as urban metrics and population size have some degree of heteroscedasticity~\cite{bettencourt2007growth}, and most of these urban indicators also follow heavy-tailed distributions~\cite{marsili1998interacting,alves2014empirical}. Thus, it is not surprising to find controversial results about the importance of variables for crime prediction when so many assumptions of linear regressions are not satisfied. 

A possible approach to overcome some of these issues is to apply a transformation to the data in order to satisfy the assumptions of linear regressions. For instance, Bettencourt {\it et al.}~\cite{bettencourt2010urban} (see also~\cite{alves2013distance,alves2015scale}) employed scaled-adjusted metrics to linearize the data and provide a fair comparison between cities with different population sizes. By considering these variables and applying corrections for heteroscedasticity~\cite{davidson1993estimation}, it is possible to describe $62\%$ of the variance of the number of homicides in function of urban metrics~\cite{alves2013distance}. Also, by the same approach, researchers have shown that simple linear models account for $31\%$--$97\%$ of the observed variance in data and correctly reproduce the average of the scale-adjusted metric~\cite{alves2015scale}. However, the data still have co-linearities which can lead to misinterpretation of the coefficients in the linear models~\cite{alves2013distance,alves2015scale}.

A better approach to crime prediction is the use of statistical learning methods (\textit{e.g.}~\cite{kang2017prediction}). Regression models based on machine learning can handle all the above-mentioned issues and are more suitable for the analysis of large complex
datasets~\cite{breiman2003statistical}. For instance, decision trees are known to require little preparation of the data when performing regression~\cite{breiman2001random,hastie2013elements,james2014introduction,de2000classification}. Tree-based approaches are also considered a non-parametric method because they make no assumption about the data. Among other advantages, these learning approaches map well non-linear relationships, usually display a good accuracy when predicting data, and are easy to interpret~\cite{breiman2001random,hastie2013elements,james2014introduction,de2000classification}. 

Here, we consider the random forest algorithm~\cite{ho1995random,amit1997shape,ho1998random,breiman2001random} to predict and quantify the importance of urban indicators for crime prediction. We use data from urban indicators of all Brazilian cities to train the model and study necessary conditions for preventing underfitting and overfitting in the model. After training the model, we show that the algorithm predicts the number of homicides in cities with an accuracy up to $97\%$ of the variance explained. Because of the high accuracy and easy interpretation of this ensemble tree model, we identify the important features for homicide prediction. Unlike simple linear models adjusted through OLS, we show that the importance of the features is stable under slight changes in the dataset and that these results can be used as a guide for crime modeling and policymakers. 

\section{Methods and Results}

\subsection{Data}
For our analysis, we choose the number of homicides at the city level as the crime indicator to be predicted. Homicide is the ultimate expression of violence against a person, and thus a reliable crime indicator because it is almost always reported. In Brazil, the report of this particular crime to the Public Health System is compulsory, and these data are aggregated at the city level and made freely available by the Department of Informatics of the Brazilian Public Health System -- DATASUS~\cite{datasus}. As possible predictor variables of crime, we select 10 urban indicators (also at the city level) available from the Brazilian National Census that took place in 2000. They are: child labor (fraction of the population aged 10 to 15 years who is working or looking for work), elderly population (citizens aged 60 years or older), female population, gross domestic product (GDP) , illiteracy (citizens aged 15 years or older who are unable to read and write at least a single ticket in the language they know), family income (average household incomes of family residents, in Brazilian currency), male population, population, sanitation (number of houses that have piped water and sewerage), and unemployment (citizens aged 16 years or older who are without working or looking for work). We have also considered as possible crime-predicting variables the number of traffic accidents and suicides (these data are also aggregated at the city level and from the same year of the census). We have chosen these indicators because they are common in studies correlating crime with socioeconomic indicators~\cite{gordon2010random}. We have further included the number of homicides in the year 2000 as a predictor for homicides in 2010 for investigating a possible autocorrelated behavior. There are thus 13 urban indicators (including also the homicide indicator) from the year 2000 in our dataset that will be used as predictors of the number of homicides 10 years later. We choose a time interval of 10 years because the characteristic time for changes in social indicators has been estimated to be of the order of decades~\cite{bettencourt2010urban,alves2015scale}. However, we also investigate how the accuracy changes as the time-lag increases from one to ten years considering homicide data from the years 2001 to 2010. 

\subsection{Problems with usual linear models}

Many statistical models try to relate crime with possible explicative variables through OLS linear regressions~\cite{gordon2010random}. Thus, it is usually assumed that crime is a linear function of the explicative variables, \textit{i.e.}, crime rate $= f($explicative variables$)$, where $f$ is a linear function~\cite{gordon2010random}. In the context of our dataset, a naive approach to model homicides in Brazilian cities is to consider the following linear regression
\begin{equation}\label{eq:linear_regression}
H(t+\Delta t) = a\, H(t) + \sum_{n=1}^{12} b_n X_n(t) + \xi(t), 
\end{equation}
where the dependent variable $H(t+\Delta t)$ is the number of homicides in the year 2010, $H(t)$ is the number of homicides in 2000, $X_n(t)$ is the $n$-th ($n=1,2,\dots,12$) urban indicator in 2000, $a$ and $b_n$ are the linear coefficients, and $\xi(t)$ is a random noise normally distributed (with zero mean and unitary variance) that accounts for unobserved determinants of homicides. It is worth noting that this model has a lagged dependent variable as an instrumental variable, which may cause problems when the error $\xi(t)$ is autocorrelated in time~\cite{wooldrige2003introductory}. 

\begin{figure*}[!ht]
\centering
\includegraphics[width=1\linewidth]{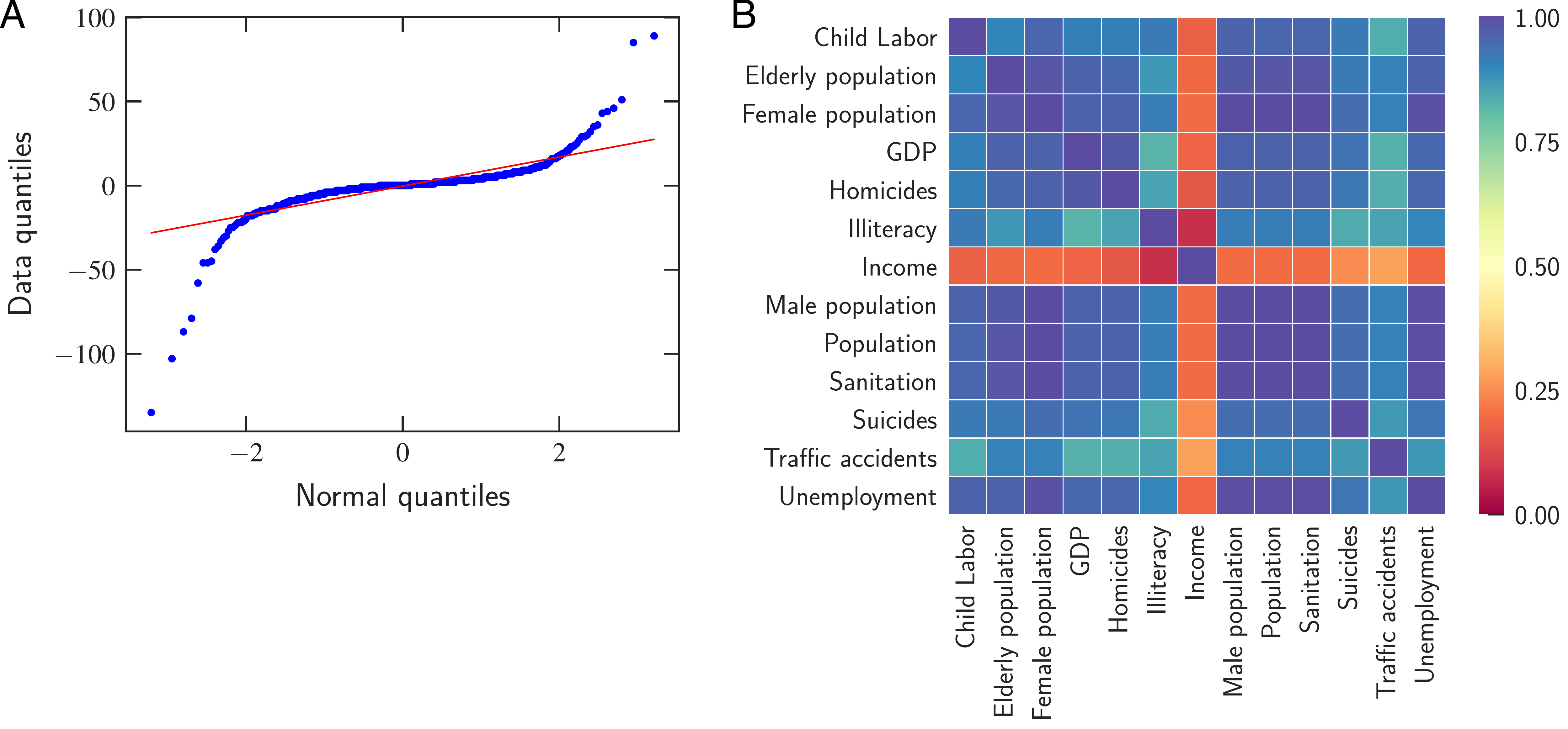}
\caption{Non-Gaussian residual and multicollinearity in urban metrics. A) Quartile-quartile plot comparing the residuals of the linear with a Gaussian distribution (represented by the straight red line). We observe systematic and huge deviations from normality, which is confirmed by Kolmogorov-Smirnov normality text ($p$-value $ < 10^{-16}$). B) The matrix of the Pearson linear correlations among the urban indicators (the predictor variables). We observe that except for income, all indicators are strongly correlated to each other, confirming the presence of multicollinearity in this dataset.}
\label{fig:dist-corr}
\end{figure*}

The OLS linear regression of Eq.~\ref{eq:linear_regression} requires the residuals to follow a normal distribution. However, this conditions is not satisfied as shown by the quantile-quantile plot in Fig.~\ref{fig:dist-corr}A, and confirmed by the Kolmogorov-Smirnov normality test ($p$-value $<10^{-16}$). Usually, researchers account for this problem by applying some transformation to data. This may eventually solve the normality problem, but collinearities may exist among urban indicators, and ignoring this fact can lead to controversial results, even when linear models display high predicting accuracies. 

According to the urban scaling hypothesis~\cite{bettencourt2007growth,bettencourt2010urban,alves2013distance,alves2015scale,leitao2016scaling,hanley2016rural}, the urban indicators in our dataset are dependent of the population size. Thus, each indicator can be written in terms of the population size $N$ as $X_n \sim N^{\alpha_n}$, where $\alpha_n$ is the urban scaling exponent, a condition that introduces correlations among all indicators. This property in regression models is called multicollinearity of the variables, and we can check for that in our data by evaluating the Pearson correlation coefficient among all urban indicators, as shown in Fig.~\ref{fig:dist-corr}B. From this correlation matrix, we verify that there are significant correlations practically between all pair of variables, violating another condition assumed by most regression models.

If we ignore the assumptions required by linear regression models and perform the OLS fitting, we could eventually find good predictions. In fact, the linear model of Eq.~\ref{eq:linear_regression} explains $87\%$ of the variance in our dataset. It is an appealing result since it is easy to interpret the coefficients of a linear model. However, the feature importance quantified by regression models can be very sensitive to changes in the dataset, such as by considering undersampling or bootstrapping methods~\cite{efron1994introduction}. For instance, by bootstrapping our dataset and applying the linear model of Eq.~\ref{eq:linear_regression}, we find that (depending on the sample used) city's income can be positively, negatively, or even uncorrelated with homicides. These facts could explain some of the inconsistencies reported in the literature about crime. For instance, Entorf {\it et al.}~\cite{entorf2000socioeconomic} found that higher income and urbanization are related to higher crime rates, whereas Fatjnzylber {\it et al.}~\cite{fajnzylber2002causes} claimed that average income is not correlated with violent crime, but higher urbanization is associated with higher robbery rates, and not with homicides. The same disagreements happen for unemployment, punishment and deterrence, among others indicators~\cite{gordon2009crime,alves2013distance,alves2015scale}. A possible way to improve the OLS regression model could be adding the variables in a stepwise fashion or considering non-linear models (e.g., the negative binomial regression) to overcome some of these problems. Unfortunately, in the crime literature, we often find results based on simple OLS linear regression, which can mislead the conclusions about the causes of crime. We could further explore and check for other inconsistencies in this simple linear regression by following the several approaches proposed in the literature of linear regression~\cite{wooldrige2003introductory}. Nevertheless, we now focus on an alternative approach to overcome these limitations and provide a meaningful rank of features under slight changes in the dataset. Our approach is much more flexible, requiring no transformation or data preparation to achieve good results regarding accuracy in predictions and meaning of results. 

\subsection{Random forest algorithm}

Random forest is an ensemble learning method used for classification or regression that fits several decision trees using various sub-samples of the dataset and aggregate their individual predictions to form a final output and reduce overfitting~\cite{breiman1996bagging,breiman2001random,hastie2013elements,james2014introduction,de2000classification}. In the random forest regression, several independent decision trees are constructed by a bootstrap sampling of the dataset. The final output is the result of a majority voting among the trees (estimators) or the average over all trees. The process of ``bagging features'' selects more often the best metrics to describe the data splits, and consequently, make them more important on the majority voting~\cite{ho1998random}. Unlike usual linear models, the random forest is invariant under scaling and various other transformations of the feature values. It is also robust to the inclusion of irrelevant features and produces very accurate predictions~\cite{hastie2013elements}. These properties of the random forest algorithm make it especially suitable for crime forecasting, because of the multicollinearity and non-linearities present in urban data.

Within the framework of random forest regression, the linear model of  Eq.~\ref{eq:linear_regression} can be rewritten as 
\begin{equation}\label{eq:random_forest}
H(t+\Delta t) =\frac{1}{B}\sum_{b=1}^B T(H(t),X_1(t),\dots, X_n(t);\Theta_b), 
\end{equation}
where $B$ is the number of trees, $T$ represents each tree adjusted to data with parameters expressed by the vector $\Theta_b$. The components of $\Theta_b$ are related to the maximum depth of the tree, splitting variables,  cutting points at each node, and the terminal-node values~\cite{breiman2003statistical}.

One common question that arises when using machine learning algorithms is related to underfitting and overfitting. These behaviors appear when estimating the best trade-off that minimizes the bias and variance errors~\cite{james2014introduction}. Bias is related to errors in the assumptions made by the algorithm that causes the model to miss relevant information about the features describing the data. In other words, the model is too simple to describe the data behavior, what is called underfitting. On the other hand, if we increase the complexity of the model by adding a great number of parameters, the model can become quite sensitive to the noise of the training set, causing an increase in the variance of the outputs, what is called overfitting. The random forest has two main parameters controlling the trade-off between bias and variance errors: the number of trees and the maximum depth of each tree. If a forest has only a few trees with a depth that goes to few layers of nodes, we are likely underfitting the data. However, if there are too many trees splitting the data in a lot of nodes to make the decisions, there is a chance that we are making choices based on the noise of the data rather than the true underlying behavior. 

\begin{figure*}[!ht]
\centering
\includegraphics[width=1\linewidth]{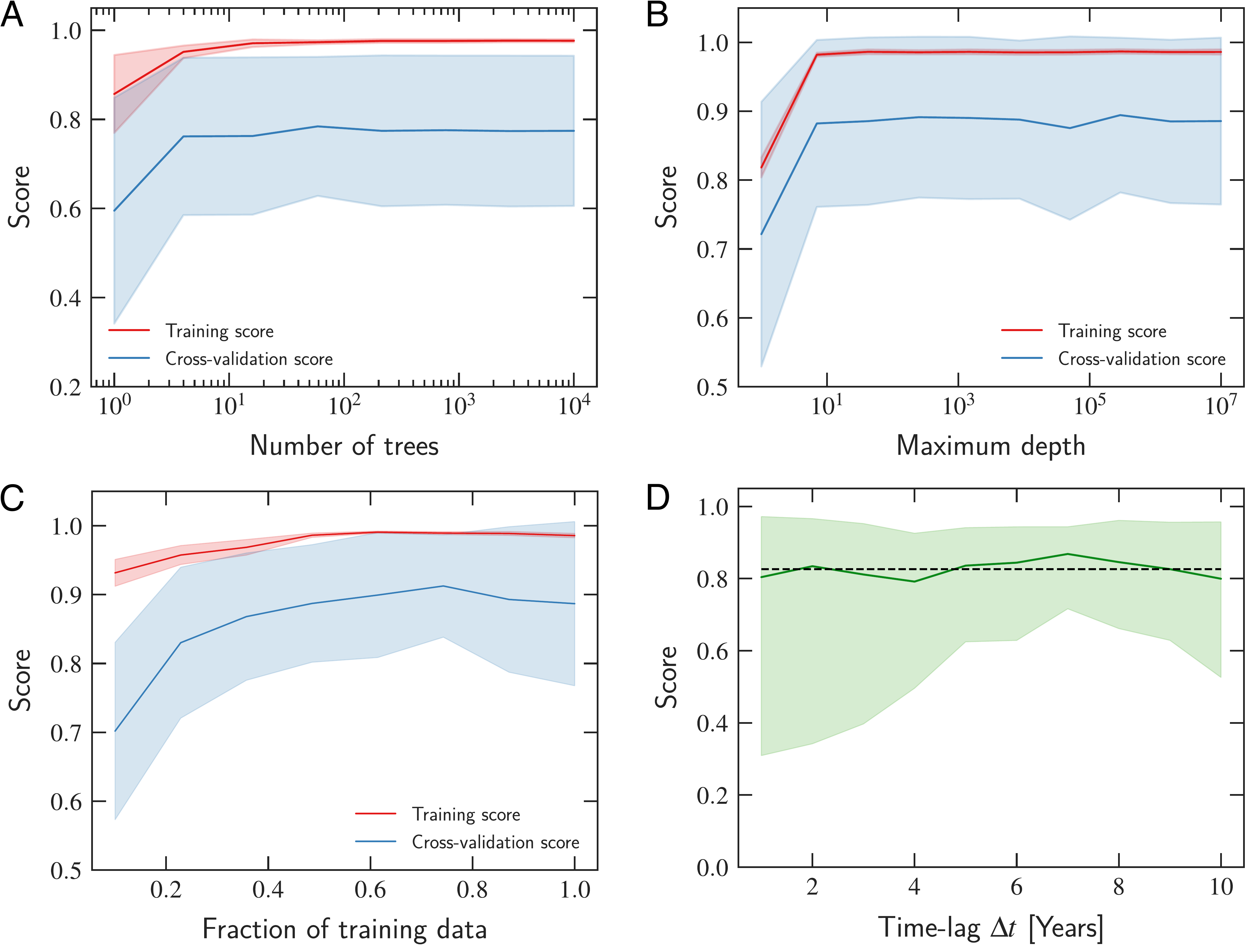}
\caption{Validation and learning curves for the random forest regressor. The solid lines represent the average score for the training set (red lines) and the cross-validated score (blue lines) for (A) different number of trees and (B) different maximum depth used in the model. For a forest with less than 10 trees and a maximum depth smaller than 10, we are underfitting the data. There is no overfitting up to $10^4$ trees and a maximum depth of $10^7$. (C) The dataset includes information of more than $5000$ cities, and the learning curve suggests that good cross-validated scores are obtained with $\approx\!20\%$ (1000 observations) of the data to train the model. We observe that the training and cross-validated scores are practically the same for data fractions larger than $\approx\!60\%$. 
In the previous plots, the shaded areas represent the confidence intervals defined by the standard deviation of the scores of the random forest algorithm. (D) The accuracy is not dependent on the time-lag $\Delta t$. The solid line shows the average score as a function of the time-lag $\Delta t$, and the shaded area stands for the $95\%$ confidence interval. The dashed line is the average score ($\approx82\%$) over all different time-lags.}
\label{fig:learning-limits}
\end{figure*}

To determine the set of parameters which avoid underfitting and overfitting of data, we use the stratified $k$-fold cross-validation~\cite{kohavi1995study,hastie2013elements} with $k=10$ for estimating the validation curves for a range of values of the parameter number of trees (Fig.~\ref{fig:learning-limits}A) and maximum depth (Fig.~\ref{fig:learning-limits}B). If the number of trees and maximum depth are both slightly smaller than 10, we find that the training score and validation score are small. As we increase the number of trees and maximum depth, both training and cross-validation scores increase and reach constant plateaus, which indicates that up to $10^4$ trees or a maximum depth equal to $10^7$ there is no overfitting. 

Another question is about how much data we need to train our model. This is a relevant question because we can eventually introduce more noise to the model by adding unnecessary data. To answer this question, we apply again the stratified $k-$fold cross-validation with $k=10$ to estimate the learning curves for a range of values of random fractions of the dataset (Fig.~\ref{fig:learning-limits}C). We observe that the more data used for training the model, the better the cross-validated scores. However, we already have good cross-validated scores with $\approx\!20\%$ of the data (about 1000 cities).

\subsection{Predicting crime with the random forest regressor}

We have to tune the model by searching for the best combination of parameters that enhances the performance of the random forest algorithm. The previous analysis of the validation and learning curves help to build a grid of parameters to seek the combination that improves our accuracy. To find the best combination of trees and the maximum depth of the trees, we use a grid-search algorithm with the stratified $k$-fold cross-validation (with $k=10$) as implemented in the Python library scikit-learn~\cite{scikitlearn}. This procedure exhaustively searches for the best combination of parameters that optimizes the score from a specified grid of parameter values. For our data, we find that the best accuracy (on average) is achieved for 200 trees and maximum depth equal 100.

\begin{figure*}[!t]
\centering
\includegraphics[width=1\linewidth]{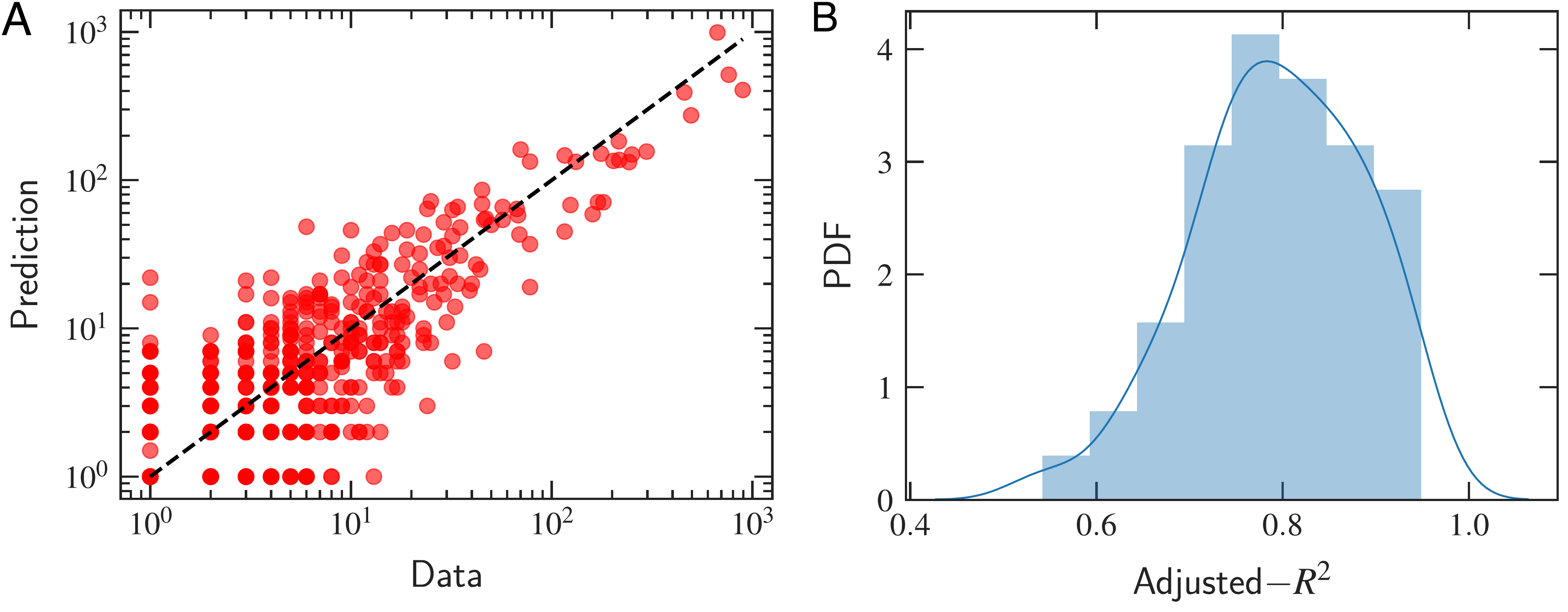}
\caption{Data versus prediction. A) The red dots are the 20\% of the data used for testing the model versus the predicted values obtained from the random forest regressor trained with 80\% of data. B) The distribution of the adjusted$-R^2$ obtained by using different random splits of the data (training and test). The boxes represent the histogram, and the solid line is the probability distribution function estimated via kernel density method.}
\label{fig:prediction}
\end{figure*}

Having the model properly trained, we can now use the random forest regressor to make crime predictions. Randomly splitting the data into $80\%$ for training and $20\%$ for testing, we obtain up to $97\%$ of accuracy in our predictions with an average adjusted$-R^2$ equal to $80\%$. Previous results using the same data and simple linear models combined with scaled-adjusted metrics were able to predict homicides with an accuracy ranging from $38\%$ to $39\%$ (see supplementary material of reference~\cite{alves2015scale}). Figure~\ref{fig:prediction}A shows the empirical data versus the random forest predictions for a realization (one data splitting) of the algorithm. Because the dataset is split randomly, different runs can return different outputs and different scores. Figure~\ref{fig:prediction}B depicts the probability distribution (computed via kernel density estimation method) of the adjusted$-R^2$ for 100 different splits, where we observe that the values are concentrated around the peak at $80\%$.

\subsection{Features importance}
The rank of features (urban indicators) importance is another intriguing question for a better understanding of crime. As previously discussed, usual linear regression models have found different answers to this question~\cite{gordon2009crime,entorf2000socioeconomic,fajnzylber2002causes}. Here, we use the random forest algorithm to identify the most important urban indicators to predict crime and test whether this ranking of features is robust under slightly changes on the dataset. 

The importance of a feature in a tree can be computed by accessing its contribution to the decision process. In each node of a single tree, the variable $X_n$ that have the best improvement in the squared error risk is used to split the region associated with that node into two subregions~\cite{hastie2013elements}. Thus, Breiman \textit{et al.}~\cite{breiman1984classification} proposed that the relative importance of a feature $X_n$ in a tree can be calculated by the sum of the improvements overall internal nodes of the tree where $X_n$ is used for splitting. The generalization of this measure to the random forest is the average importance of the features over all trees in the model, as implemented in the Python library scikit-learn~\cite{scikitlearn}.

\begin{figure*}[!ht]
\centering
\includegraphics[width=1\linewidth]{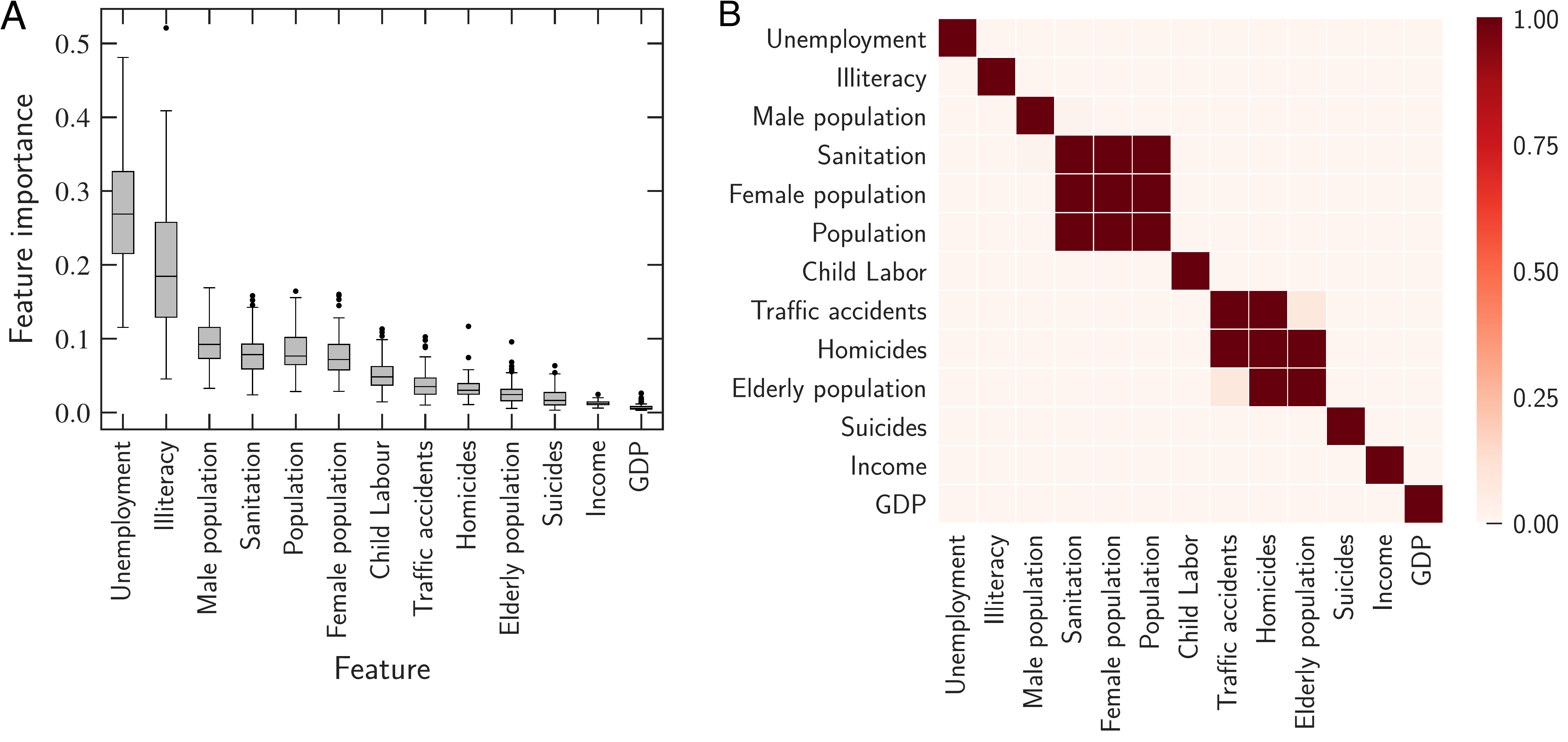}
\caption{The important urban metrics for predicting the crime of homicide. A) The box-plot shows the rank the importance of urban metrics by their medians over different random sample splits. The black lines dividing the boxes represent the median, boxes represent the interquartile range, upper and lower whiskers bars are the most extreme non-outlier data points, and dots are the outliers. B) The matrix plot shows the $p-$values of $t-$Student test comparing whether the positions in the rank are different. We identify four groups of features that are equally important in the rank (squares on the diagonal matrix).}
\label{fig:importance}
\end{figure*}

We use this metric to calculate the importance of the features describing the number of homicides in our data. Because of the slight modifications in the dataset caused by the splits in training and test samples, the feature importance varies for each realization of the algorithm. To verify whether these changes affect the importance rank of the features, we calculate the importance of the features for 100 different samples and compute a box-plot ranked by the median importance of the outputs returned by the different samples (Fig.~\ref{fig:importance}A). We check whether the differences in importance are statistically significant by computing the $p-$values of the $t-$Student test (testing if two samples have identical average values) with Bonferroni corrections (that corrects the $p-$values for multiple comparisons~\cite{rupert2012simultaneous}). The resulting $p-$values are shown in Fig.~\ref{fig:importance}B and ranked by the median in correspondence with the box-plot. From this figure, we note that unemployment is the most important feature to describe the crime, followed by illiteracy and male population. The next urban indicators in order of importance form groups of indistinguishable importance (squares on the diagonal matrix of Fig.~\ref{fig:importance}B), \textit{i.e.}, the fourth most important feature is actually a group that includes female population, population, and sanitation. This group of features is followed by other three groups: child labor and homicides (at the fifth position), traffic accidents and elderly population (at the sixth position), and suicides and income (at the seventh position). The less important feature to predict crime is the GDP of cities. It is interesting to note that the number of homicides in past is a poor predictor of homicides in future. It is also worth remarking that despite some fluctuations in the rank caused by the different samples used to train the model, the ranking of features remains the same. This result demonstrates the robustness of the random forest algorithm for detecting the importance of features.

\section{Discussion and conclusions}

The accurate predictions obtained through statistical learning suggest that crime is quite dependent on urban indicators. The easy interpretation and good accuracies of the random forest algorithm show that this model is an excellent solution for predicting crime and identify the importance of features, even under small perturbations on the training dataset. We cannot assert from the rank of Fig.~\ref{fig:importance}A which features have a positive and negative contribution to the number of homicides. We could try to decompose the contributions of each feature in each node of the trees, and calculate an average over all nodes and trees to identify the gradients, as described by Hastie {\it et al.}~\cite{hastie2013elements}. However, this decomposition is problematic because features can contribute differently depending on thresholds imposed by the constraints of the sample chosen to train the model. The signals of the contributions of a particular feature can vary for different thresholds, even when the average rank of importance remains the same~\cite{breiman1984classification}. Thus, despite the great stability of the results, it is hard to identify whether the variables cause a positive or a negative effect on the number of homicides. While one can speculate whether the most important indicators found in our results cause crime to increase or decrease, further investigation is necessary to understand the local effects of urban metrics on crime.

Because unemployment has a very large variability at local level~\cite{levitt2001alternative}, Blumstein~\cite{blumstein2002crime} argue that different data aggregation can lead to different conclusions on whether this indicator affects crime or not. There is also evidence supporting the idea that crime is affected by unemployment only when this indicator exceeds a given threshold~\cite{alves2013distance}. This is somehow similar to what random forest does when separating the hyperplane of features by thresholds and classifying the number of homicides according to the different values of the urban indicators. The algorithm cannot indicate whether unemployment contributes positively to crime; however, it indicates that this indicator is the most important for describing crime among the set of 12 features in our dataset. Particularly, a recent work has shown that the raising of shooting in schools is related to unemployment rate across different geographic aggregation levels (national, regional and city)~\cite{pah2017economic}, which is in agreement with our findings.

The second most important feature for describing crime is illiteracy, which has been associated with violence by other works~\cite{davis1999low}. Previous works on scaled-adjusted metrics have also shown that the levels of illiteracy are correlated to the number of homicides~\cite{alves2013distance,alves2015scale}. A report made by the Canadian Police further shows that people with low literacy skills are less likely to involve in group activities than those with higher literacy skills~\cite{reportpolice}. Consequently, low literate people often feel isolated and vulnerable, making them more likely to involve in violence and crime~\cite{reportpolice}. The male population is the third most important feature for describing homicides and has also been linked to high levels of violence~\cite{hesketh2006abnormal,alves2013distance,alves2015scale}. As discussed by Hesketh and Xing~\cite{hesketh2006abnormal}, the surplus of male population increases the marginalization in society and is linked to antisocial behavior and violence. A similar result was reported in Ref.~\cite{alves2013distance}, where it was found that cities with more man than the expected (in terms of the urban scaling) usually have more crimes. Similarly, our results suggest that male population plays an important role in defining the levels of crime in cities. Notice that together unemployment, illiteracy, and male population are responsible for explaining $78\%$ of the variance in our dataset.

It is worth mentioning that previous results based on the same data and simple linear models combined with scaled-adjusted metrics were able to predict homicides with accuracy ranging from $38\%$ to $39\%$ (see supplementary material of reference~\cite{alves2015scale}), a much lower score compared to the ones obtained with the random forest approach. However, more sophisticated methods could improve even more this accuracy, and this could be the target of future investigations on crime. The literature of crime still lacks a full comparison with other methods (such as non-linear models, Bayesian regression, and other machine learning techniques), a work that can unveil new properties of the data as well as improve the accuracy of predictions.

Finally, we believe that the application of machine learning for identifying urban indicators that correlate with crime helps to settle the discussion about whether an indicator is important or not for describing a particular crime type. The results of our analysis can further be used as a guide for building other crime models and may help policymakers in the seek of better strategies for reducing crime. Indeed, our results indicate that unemployment and illiteracy levels play an important role in defining the number of homicides in Brazil. 

\section*{Acknowledgments}
L.G.A.A. acknowledges FAPESP (Grant No. 2016/16987-7) for financial support. H.V.R. acknowledges CNPq (Grant No.  440650/2014-3) for financial support. F.A.R. acknowledges CNPq (Grant No.  307748/2016-2) and FAPESP (Grant No. 2016/25682-5 and Grant No. 13/07375-0) for financial support.

\end{document}